# Quantized vortex rings in superfluid He$^4$ and Light Dark Matter

Massimo Cerdonio*
INFN Section and University, Padova, Italy
*cerdonio@pd.infn.it

Abstract
The possibility is considered that, after nuclear recoil caused by elastic scattering with Light Dark Matter, atoms in superfluid He$^4$ may create quantized vortex rings, which can thereafter be detected. The concept for sort of a "telescope" for LDM is presented, which would be sensitive to its motion through the Galactic LDM halo. A demonstrator is proposed, which makes use of thermal neutrons.

Introduction

Direct searches of Dark Matter, DM, rely on recoil methods as the DM, due to its local dispersion velocity in the Galactic halo, imparts a kick to ordinary matter. A variety of materials and methods are in use to detect such recoils [1]. Since superfluid He$^4$ was proposed long ago for measurement of pp solar neutrino flux – the HERON project [2] – this field evolved towards proposals for searches of DM, with the aim to push the direct detection parameter space to masses below 1 $Gev/c^2$ [3]. Quite recently a proposal appeared to push the lightest detectable DM into the $keV/c^2$ region [4].

An aspect of the processes, which can take place after nuclear recoil, is that a He$^4$ atom is put in "fast" motion within the superfluid. This situation shows a few peculiarities, which are due to the quantum nature of the superfluid, and at the same time to its behavior as a perfect fluid of classical hydrodynamics. We elaborate here on this matter to come up with a proposal for a Light Dark Matter, LDM, detector based on those properties of superfluid He$^4$.

The Landau criterion for superfluidity [5] connects, via conservation of total energy and momentum, an initial state of the prefect fluid flow pattern in presence of an obstacle moving at a relative velocity $v$, with a final state giving a different flow pattern plus excitation(s). The criterion entails the notion of a *critical velocity* $v_c$, above which the transition process can occur – a selection rule. In other words, with an example relevant here, if a DM particle kicks a He$^4$ nucleus and puts in motion its atom in the superfluid, the breakdown of superfluidity and creation of an excitation will be allowed only if the velocity $v_{He}$, which the atom would gain, can be $v_{He} > v_c$. The appropriate critical velocity $v_c$ will be dictated by the Landau criterion, applied to whichever the lowest lying branch of excitations is.

A branch of excitations, which thermodynamically is found to exist – the density of such excitation makes up for the "normal" component in the "two fluids" model [5] - and which has been directly probed by thermal neutrons, is commonly described as dominated by "phonons"- acoustic modes in the fluid - and "rotons"- modes of the Feynman and Cohen "backflow" motion of the fluid around fast atoms

[6]. The critical velocities for creation of phonons and rotons are called "Landau critical velocities" and are respectively about 240 m/s – the velocity of sound – and 60 m/s.

Another branch of excitations exists, which can satisfy the Landau criterion at lower velocities – the Onsager-Feynman quantized vorticity [6]. In particular electrons and positive ions injected in liquid He$^4$ behave as hard sphere - for a brief introduction, see Chap 1 in ref [7] – and thus are ideal probes to look at creation of such lower lying excitations. In fact the classic experiment of Rayfield and Reif [8] showed how they both prefer to create *quantized vortex rings*. Furthermore the ions get trapped in the vortex core, so that the vortex can be moved around with electric fields and detected from the charge it carries. In brief a consistent picture emerged [9]: the vortex ring, at temperatures below some 300 mK, moves undisturbed by phonons/rotons, and consists of the superfluid flowing as a perfect fluid around a filamentary ring of width of order Å and diameter of order hundreds Å. The circulation is quantized in units of $h/m_{He}$, with $h$ the Planck constant and $m_{He}$ the bare He$^4$ atomic mass – an astonishing medley of classical and quantum physics.

These introductory remarks stimulate us to consider a possibility complementary to that advanced in ref [4]. Should an LDM particle have an elastic scattering with a helium nucleus and thus kick an He$^4$ atom in the superfluid, we consider if the final result can be the creation of a quantized vortex ring, the detection of which would give a detection of the incoming LDM particle. We find the process allowed by the Landau criterion, and give here the order of magnitude of the lighter LDM, which would be detectable, possibly about 100 keV/c$^2$.

*The concept*
In refs [8,9] comparatively large rings were observed and the question of the critical velocities needed for the creation process was not addressed. Schwarz and Jang [10] developed a model, which came to be the closest to give agreement with a wealth of experimental situations – see discussion in [10] and in [7]. Here we build on ref [10] for an estimate of the effect of nuclear recoil of an LDM particle in superfluid He$^4$.

In liquid He$^4$, and in particular in the superfluid phase, an excess electron self-traps in a spherical bubble of radius ~ 17 Å, while an excess positive charge gets surrounded by a snowball of solid helium of radius ~ 5 Å. In both cases the boundary with the liquid is sharp ~ 1 Å [10 and refs therein]. So the Authors in [10] are led to consider a hard sphere of given radius and mass moving with velocity $U$ in the superfluid, and look for the minimum $U_c$, which gives the creation of a quantized vortex ring while the sphere is slowed down. This is done by extending the Landau criterion by taking in full account the classical hydrodynamics of a perfect fluid. One crucial issue is to realize that a vortex in the wake of the sphere must be accompanied by a counter circulating image to fulfill the boundary conditions at the sphere-liquid interface. It is found that the minimum velocity to satisfy the (extended) Landau criterion is the one which corresponds to the vortex just

"girdling" the hard sphere. In such a situation, it happens that, quoting from ref [10], "In this configuration...the velocity fields of the real and the image ring cancel most effectively, so that the energy of the liquid flow is particularly small." This is what makes possible to obey the Landau criterion with a *single excitation* process at energies and momentum transfers lower than those needed for the excitation of single phonons/rotons. The theory well reproduces qualitatively a wealth of experimental findings. Quantitatively it appears to work up to few percent for the bigger ions (the negative), while for the smaller ones (the positive) the agreement is only within many tens of percent.

Now we can formulate our concept. A "fast" atom can be considered an impenetrable sphere of radius between 1.4 Å – its van der Waals radius - and 1.9 Å, depending if its effective mass is the bare mass or 2.3 times the bare mass respectively [6 and refs therein]. It is only a factor of 3 smaller than that of a positive ion. In the spirit of ref [6] let us see what would happen to a helium atom, when it is made to move "fast" within the fluid, but not fast enough to create a phonon/roton. We may try to guess from [10] what it would be the critical velocity for creation of a vortex, which girdles spheres of 1.4 Å - 1.9 Å radiuses. From the overall contest of the calculations of ref [10], it is plausible that such a critical velocity may stay at the order of 10 m/s.

Unfortunately it is also clear from ref [10] that the smaller the sphere radius the more the simplifying assumptions, which give O(1) agreement for a radius ~ 5 Å, will induce larger errors. In fact the approximation that the system continues to behave as a perfect fluid down to the atomic scale must fail as this scale is approached. It is comforting that the critical velocity for creation of a vortex ring remains in the same ball park as that of the ions, giving thus plausibility to our proposal, but we cannot ask more from theoretical considerations and so something else has to be done to progress (see below).

Meanwhile let us better focus the concept of detection. If an LDM particle, traveling at the local dispersion velocity $v_X$ ~ 0.8 $10^{-3}$c [11] kicks a He nucleus, and in the recoil the He atom acquires a velocity of order $v_{He}$ ~ 10 m/s relative to the superfluid, such a "fast" atom on one hand will not have the possibility to create a single phonon/roton, because its velocity would be below critical for that process, but on the other hand may have the critical velocity for creation of a quantized vortex ring. For a LDM mass $m_X$ << $m_{He}$ – taking $m_{He}$ ~ 4 GeV/$c^2$ - the maximum momentum transfer in a head-on elastic scattering would be 2$m_X$ $v_X$ ~ $v_{He}$ $m_{He}$. Thus the creation of a quantized vortex ring will be allowed if $m_X$ is of the order $m_X$ ~ 100 keV/$c^2$.

A conceptual scheme for an experiment would be like this. The LDM enters a long cylindrical tube of superfluid He$^4$ kept at, say, temperatures of order of 100 mK. At these temperatures and lower the density of the "normal" component – phonons and rotons – is so low that the hard sphere would feel negligible viscosity, and thus can be described as an object moving freely in a perfect fluid. In the initial region the LDM interacts with a He$^4$ atom, which creates a vortex ring, which in turn continues to drift towards a detection region, where a cloud of negative ions has been created. Here the (small) vortex gets attached to the (large) "hard sphere" [12] – the negative ion – and, with appropriate electric fields, can acquire energies of order eV. Finally such an energetic vortex can be detected in a number of ways. One is to let it annihilate on the terminal wall of the cylinder, where it is detected thermally, as it releases energy to transition edge sensors, TES. Alternatively, again dragged by appropriate electric fields, can be brought to annihilation on a section of free surface, where a single electron can be extracted and counted, as demonstrated in [13]. So in any case it appears possible to detect a single vortex, and thus a single LDM.

Notice that, for a large aspect ratio of the cylinder, such a set-up makes for sort of "telescope" for the LDM, as the maximal response will happen when its axis is aligned with the direction of the incoming LDM. Therefore, in case of a number of detections, the diurnal and seasonal distribution will give a relevant signature. The kinematics of the hard sphere elastic scattering will elucidate the LDM mass range which is detectable, once the geometry of the detection region is fixed and gives the acceptance angle for non-head-on scatterings [14]. Notice the attractive of this concept in that the detection process ultimately involves detection of eV energies in relation to a single LDM recoil event, rather than the meV energy sensitivities required for detection of phonons/rotons, as for the concept given in [4].

The always present remnant vorticity [15], which develops on cooling through the lambda transition, should not interfere with the above scheme. Is is believed that the small ions and associated small vortex rings do not interact significantly with the remnant vorticity [7]. This should be even truer for the smaller vortex ring associated with our "fast" atom.

*Proposal for a demonstrator*
The method outlined above would be somewhat complementary to that of ref [4]. Actually both processes may occur, and thus compete, in the appropriate mass range. That of ref [4] is a second order process, while the present one appears to be at lower order, and thus preferred, but it is hard to predict what could be the branching ratio. In fact a theoretical investigation looks as an overwhelming task, due primarily to the complexity of the detailed behavior of ions in liquid helium, together with the multitude of contrasting interpretations, which have been advanced - see a wealth of discussions in ref [7].

Thus one may consider a direct experimental approach. Nuclear recoil, similar to the LDM interaction, apart for the much larger cross-section, can be provided by elastic scattering of slow neutrons. Then the set up for a demonstrator would be quite similar to the LDM "telescope" described above.

A beam of neutrons of selected velocity would enter from one side of the demonstrator, kick $He^4$ atoms and create quantized vortex rings, which would be detected as above. So this test of the system would become sort of calibration of the detector, once the flux and the cross section of the neutrons is taken in account. It should be no show stopper the fact that, apart one clearly presented instance [16], no attempt is found to detect inelastic neutron scatterings, which would directly create a vortex ring. In fact one has to consider in this respect that the process of vortex creation by an ion is pretty slow [17,18]. Thus the neutron must scatter *elastically* from the helium nucleus and put in motion its atom. Only later on, on a time scale separated from that of the impact with the neutron, the "fast" moving helium atom is ultimately inelastically slowed down by the creation of the quantized vortex ring.

Finally an experiment with such a demonstrator appears to be feasible with on the shelf technologies.

*Conclusions*

We offered a plausibility argument for a process of creation of quantized vortex rings in pure superfluid He$^4$ by Galactic LDM. As a theoretical analysis appears quite difficult, to understand if such a process actually occurs and what are its features, we propose a demonstration based on elastic scattering of slow neutrons. If the process is actually observed, results from the demonstrator would assess what are the critical velocities for creation of quantized vortex rings after nuclear recoil with thermal neutrons and give the corresponding effective cross section.

If it happens that the critical velocities are actually pretty low, then one would be able to detect/limit LDM masses $m_X$ in the 100 keV range. In this case one could predict completely the performance of an actual detector, and see how deeply the search would invade this unexplored region of parameter space.

In any case the experiments with the demonstrator would be relevant *per se* for the physics of the breakdown of superfluidity.

*Acknowledgements*

This paper is dedicated to the late Professor Giorgio Careri, a pioneer of investigating superfluid vorticity in He$^4$ with ions, and the Master who introduced the author to the field as a student in early 60'. Extensive and fruitful discussions with Giovanni Carugno and Antonello Ortolan are much gratefully acknowledged.